# Three new infrared bands of the He - OCS complex

J. Norooz Oliaee, [1] B.L. Brockelbank, [1] A.R.W. McKellar, [2] N. Moazzen-Ahmadi 1*

[1] *Department of Physics and Astronomy, University of Calgary, 2500 University Drive North West, Calgary, Alberta T2N 1N4, Canada.*

[2] *National Research Council of Canada, Ottawa, Ontario K1A 0R6, Canada.*

**Abstract**

Three new infrared bands of the weakly-bound He-OCS complex are studied, using tunable lasers to probe a pulsed supersonic slit jet expansion. They correspond to the $(04^00) \leftarrow (00^00)$, $(10^01) \leftarrow (00^00)$, and $(04^01) \leftarrow (00^00)$ transitions of OCS at 2105, 2918, and 2937 cm$^{-1}$, respectively. The latter band is about 7900 times weaker than the previously studied OCS $\nu_1$ fundamental. Vibrational shifts relative to the free OCS monomer are found to be additive. Since carbonyl sulfide has previously been shown to be a valuable probe of superfluid quantum solvation effects in helium clusters and droplets, the present results could be useful for future studies of vibrational effects in such systems.

Address for correspondence:   Prof. N. Moazzen-Ahmadi,
                              Department of Physics and Astronomy,
                              University of Calgary,
                              2500 University Drive North West,
                              Calgary, Alberta T2N 1N4,
                              Canada.

* Corresponding author. Tel: 1-403-220-5394

## 1. Introduction

Intermolecular interactions between He and OCS are of basic interest, but also have taken on practical importance because OCS turns out to be an especially useful probe of microscopic superfluid effects in helium clusters and droplets. This has led to numerous studies of "quantum solvation" in helium probed by OCS, both experimental [1 - 9] and theoretical [11 - 16]. Detailed intermolecular interaction potentials have been calculated for He-OCS, with increasing accuracy over the years [17 - 24]. These potentials can be rigorously tested by comparison with experimental bound-state energy levels of the He-OCS van der Waals complex derived from high-resolution spectra, as observed in the microwave [17, 25] and infrared [26 - 28] regions.

Previous infrared studies of binary He-OCS complexes, and of $He_N$-OCS clusters, have been limited to the strong $\nu_1$ fundamental (C-O stretch) band of OCS located at 2062 cm$^{-1}$. In the present paper, we extend the infrared study of He-OCS to three OCS combination bands at 2105, 2918, and 2937 cm$^{-1}$. Though these transitions are much weaker than $\nu_1$, they already provide new information on vibrational shift effects, and could prove useful for future studies of vibrational effects in superfluid helium clusters.

The He-OCS complex has a T-shaped structure, with a binding energy ($D_0$) of about 19 cm$^{-1}$ [23], a center of mass separation of about 3.83 Å, and an effective angle of about 66° between the O-C-S molecular axis and the line connecting the centers of mass [6]. It is thus a prolate asymmetric rotor with the *a*- and *b*-inertial axes approximately aligned with the O-C-S and He-C axes, respectively, and the *c*-axis perpendicular to the plane. Since the permanent dipole moment, and stretching transition moments, lie along the O-C-S axis, observed microwave and infrared transitions are predominately $a$-type ($\Delta K_a = 0$, $\Delta K_c = \pm 1$) with a weaker $b$-type ($\Delta K_a = \pm 1$, $\Delta K_c = \pm 1$) contribution .

- **2. Results**

Spectra were recorded at the University of Calgary using a pulsed supersonic slit jet expansion probed by a tunable infrared diode laser (for 2105 cm$^{-1}$) or optical parametric oscillator source (for 2918 and 2937 cm$^{-1}$) as described previously [28-31]. A typical expansion mixture contained 0.1 - 0.2% carbonyl sulfide in helium carrier gas with a jet backing pressure of 10 - 17 atmospheres. Spectral assignment and simulation were made using the PGOPHER software.[32] We label the OCS vibrational states using ($v_1$, $v_2^{l2}$, $v_3$), where $v_1$ represents the C-O stretch, $v_2$ the bend (with angular momentum, $l_2$), and $v_3$ the C-S stretch, with fundamental values of 2062.2, 520.4, and 859.0 cm$^{-1}$ for $^{16}O^{12}C^{32}S$. The He-OCS band in the region of the $\nu_1$ fundamental band has been studied previously in detail [26 - 28], including some isotopically substituted forms ($^3$He, $^{34}$S, $^{13}$C) in addition to the main isotopologue, $^4$He-$^{16}O^{12}C^{32}S$.

The OCS combination band $(04^00) \leftarrow (00^00)$ is located at 2104.828 cm$^{-1}$. This is 42.627 cm$^{-1}$ above the $\nu_1$ fundamental band, from which it "steals" some intensity through anharmonic (Fermi) interaction. As a result, the $(04^00) \leftarrow (00^00)$ band is much stronger than it would otherwise be, though still about 250 times weaker than $(10^00) \leftarrow (00^00)$ [33 - 35]. We have now observed the spectrum of He-OCS in this region, as shown in the top panel of Fig. 1. Rotational analysis of this band was relatively easy, and we assigned a total of 60 transitions with values of $J'$ and $K_a'$ up to 6 and 3, respectively. For comparison, in the fundamental band it was possible to observe 133 transitions up to $(J', K_a') = (8, 4)$ [28].

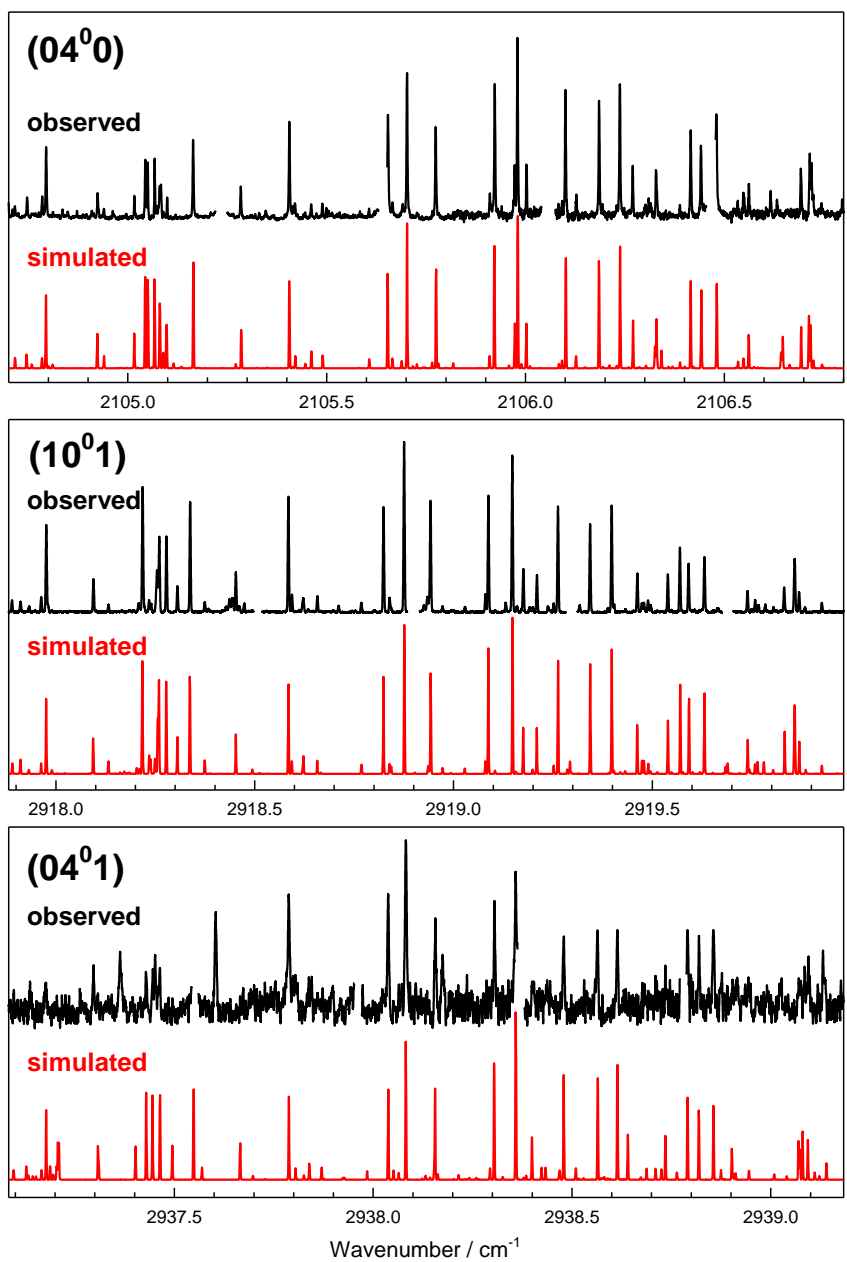

Figure 1. Observed and simulated spectra of He-OCS accompanying three combination bands of OCS. Gaps in the experimental traces are regions obscured by OCS monomer transitions.

Table 1. Molecular parameters for He - OCS complexes (in cm$^{-1}$).[a]

|  | $(00^00)$ | $(10^00)$ | $(04^00)$ | $(10^01)$ | $(04^01)$ |
|---|---|---|---|---|---|
| $\nu_0$ |  | 2062.3125(2) | 2105.0997(4) | 2918.2810(3) | 2937.4832(4) |
| $\Delta\nu_0$ |  | +0.1117 | +0.2720 | +0.1761 | +0.3364 |
| $A$ | 0.440779(3) | 0.43622(15) | 0.45326(68) | 0.43818(21) | 0.45134(54) |
| $B$ | 0.18337(4) | 0.18304(9) | 0.18731(14) | 0.182562(48) | 0.18120(26) |
| $C$ | 0.12236(4) | 0.12136(9) | 0.11928(15) | 0.121066(38) | 0.12326(26) |
| $10^3 \times \Delta_K$ | 1.694(5) | 0.94(3) | 10.47(20) | 1.680(35) | 5.40(11) |
| $10^5 \times \Delta_{JK}$ | 3.97(18) | 5.6(12) | 10.5(66) | 2.98(104) | 45.0(62) |
| $10^5 \times \Delta_J$ | 3.105(11) | 2.97(9) | 9.65(23) | 3.247(60) | -2.79(86) |
| $10^4 \times \delta_K$ | 1.96(21) | 4.3(5) | 16.31(49) | 4.36(12) | -3.6(13) |
| $10^5 \times \delta_J$ | 1.086(3) | 1.07(7) | 3.24(10) | 1.244(34) |  |
| $10^4 \times H_K$ | 0.270(15) | -1.8(3) | 7.91(12) | 0.400(15) |  |
| $10^5 \times H_{KJ}$ | 2.05(13) | 0.5(3) | 4.99(99) | 0.397(97) |  |
| $10^5 \times H_{JK}$ | -0.35(3) | 0.07(8) | 0.56(13) |  |  |
| $10^7 \times H_J$ | -0.25(4) | -0.37(12) |  |  |  |
| $10^5 \times h_K$ | -1.9(3) | 1.5(7) |  | 1.42(12) |  |
| $10^7 \times h_J$ | -0.46(3) | -0.21(11) |  |  |  |
| rms error |  | 0.0008 | 0.00085 | 0.00068 | 0.00057 |

[a] Uncertainties in parentheses are $1\sigma$ from the least-squares fits, expressed in units of the last quoted digit. The $(00^00)$ and $(10^00)$ parameters are from [28]; remaining parameters are the current work.

The $\nu_1 + \nu_3$ combination band of OCS, located at 2918.105 cm$^{-1}$, is about 70 times weaker than $\nu_1$, which makes it about 3.5 times stronger than the $4\nu_2$ band just described. Our observed spectrum of He-OCS in this region is shown in the middle panel of Fig. 1. Here, we assigned 114 transitions up to $(J', K_a') = (8, 4)$.

Just as the $(04^00)$ state steals intensity from $(10^00)$, so $(04^01)$ also steals from $(10^01)$. However, in the latter case the separation of the two states is only 19 cm$^{-1}$ (as compared to 43 cm$^{-1}$), so the interaction is somewhat stronger, and the $(04^01) \leftarrow (00^00)$ band turns out to be about 110 times weaker than $(10^01) \leftarrow (00^00)$, and hence about 7900 times weaker than $\nu_1$ [33]. Our observed spectrum of He-OCS in this region is shown in the bottom panel of Fig. 1. It is immediately evident that the signal to noise ratio in this spectrum is much worse than the others, which is not surprising given the weakness of the OCS band. Detection of this band was possible by implementation of quantum-correlated twin beams (idler and signal) for cancellation of the power fluctuations in the rapid-scan mode [36]. Even so, we were able to assign 23 transitions up to $(J', K_a') = (5, 2)$ and these assignments were virtually unambiguous, thanks to the relatively simple and widely-spaced nature of the He-OCS spectrum.

We fitted the observed transitions using the Watson A-reduced Hamiltonian with results as shown in Table 1. Ground state parameters were fixed at those determined previously [28] using microwave [17, 25] and $\nu_1$ fundamental band [28] data. The table includes these ground and $\nu_1$ state parameters, for comparison with the new results. Complete lists of observed and calculated line positions are given in Tables A1-A3. The root mean square errors of the fits were 0.00085, 0.00068, and 0.00057 cm$^{-1}$ for the three new bands, $(04^00)$, $(10^01)$, and $(04^01)$, respectively. These values are similar to that achieved for the $\nu_1$ fundamental [28], but larger than the estimated relative experimental accuracy which is about 0.0002 to 0.0003 cm$^{-1}$. This may be a reflection of the inadequacy of a conventional (semi-rigid molecule) Hamiltonian to fit the rotational levels of a floppy and weakly-bound complex such as He-OCS, in spite of including many centrifugal distortion parameters. Similarly, note that the ratios of the number of fitted (upper state) levels to the number of varied parameters were relatively small: 28/13, 49/13, and 18/9 for the three bands. For these reasons, one should be cautious about the significance of the parameters in Table 1 (especially the higher-order ones) and be aware that their quoted standard deviations likely underestimate the true uncertainties.

- **3. Discussion and conclusions**

It is interesting to consider the He-OCS band origins in terms of their shift, $\Delta\nu_0$, relative to the free OCS molecule, as summarized near the top of Table 1. In all cases, there is a small positive ("blue") shift (0.11 to 0.34 cm$^{-1}$), indicating that the van der Waals bond becomes slightly weaker in the excited vibrational states. The shifts observed here for the combination bands are larger than that of the $\nu_1$ fundamental. Most notable is the fact that the shift for the ($04^01$) upper state is exactly equal (within experimental error) to the sum of the shifts for ($04^00$) and ($10^01$) minus the shift for ($10^00$). This is just what one would expect if the shifts were simple additive functions of the OCS vibrational quantum numbers. In this context, it would be interesting to measure the He-OCS shift for the OCS $\nu_3$ fundamental band at 859 cm$^{-1}$, which should be equal to +0.0644 cm$^{-1}$ if this trend continued.

Experimental line widths in the current spectra ($\approx$0.003 cm$^{-1}$) were limited by residual Doppler broadening due to the (multi-passed) laser beam not being perfectly orthogonal to the velocities of the molecules in the supersonic expansion. There was no evidence of further broadening due to predissociation, that is, due to finite upper state lifetimes. The present results show that it is now possible to detect spectra of weakly-bound complexes even for fairly weak infrared bands. In particular, the current result for the $(04^01) \leftarrow (00^00)$ band of OCS (transition dipole moment $\approx 0.0036$ Debye [33]) probably represents the weakest of many such bands studied using our apparatus over the past ten years.

**Acknowledgments**

Financial support from the Natural Sciences and Engineering Research Council of Canada is gratefully acknowledged. We thank A.J. Barclay and K. Esteki for assistance with the experiment.

**Appendix A. Supplementary material**

Supplementary data associated with this article can be found, in the online version at http://dx.doi.org/10.1016/j.jms.2017.xx.xxx.

Appendix A

Table A1. Observed and calculated transitions in the (0400) <-- (0000), band of He-OCS dimer around 2105 cm$^{-1}$ (units of cm$^{-1}$).

```
************************************************************************
   J'   Ka'  Kc'    J"   Ka"  Kc"   Observed    Calculated    Obs-Calc
************************************************************************
    0    0    0     1    0    1    2104.79448   2104.79406    0.00041
    0    0    0     1    1    1    2104.53816   2104.53790    0.00025
    1    1    0     2    1    1    2104.43374   2104.43479   -0.00105
    1    1    0     1    1    1    2105.16416   2105.16492   -0.00076
    1    1    0     1    0    1    2105.42031   2105.42108   -0.00077
    1    0    1     2    0    2    2104.49886   2104.49887   -0.00001
    1    0    1     0    0    0    2105.40609   2105.40587    0.00021
    1    0    1     1    1    0    2104.78455   2104.78404    0.00050
    1    1    1     2    1    2    2104.55316   2104.55320   -0.00004
    1    1    1     1    1    0    2105.04400   2105.04362    0.00037
    1    1    1     0    0    0    2105.66549   2105.66545    0.00003
    2    0    2     1    0    1    2105.70206   2105.70231   -0.00025
    2    0    2     2    1    1    2104.71582   2104.71602   -0.00020
    2    1    2     1    0    1    2105.91034   2105.90944    0.00089
    2    1    2     2    1    1    2104.92392   2104.92314    0.00077
    2    1    2     3    1    3    2104.28394   2104.28296    0.00097
    2    1    2     1    1    1    2105.65414   2105.65327    0.00086
    2    2    0     1    1    1    2106.54866   2106.54829    0.00036
    2    2    0     2    2    1    2105.06737   2105.06721    0.00015
    2    1    1     1    1    0    2105.77418   2105.77538   -0.00120
    2    1    1     2    1    2    2105.28401   2105.28496   -0.00095
    2    1    1     2    0    2    2105.48915   2105.49020   -0.00105
    2    2    1     2    2    0    2105.04995   2105.05009   -0.00014
    3    1    2     3    0    3    2105.60652   2105.60714   -0.00062
    3    1    2     2    1    1    2106.10054   2106.10155   -0.00101
    3    1    2     3    1    3    2105.46111   2105.46137   -0.00026
    3    2    2     3    2    1    2105.01643   2105.01635    0.00007
    3    2    2     2    2    1    2105.97275   2105.97321   -0.00046
    3    2    2     3    1    3    2106.08291   2106.08399   -0.00108
    3    3    0     2    2    1    2107.35775   2107.35732    0.00042
    3    3    0     3    3    1    2105.08090   2105.08102   -0.00012
    3    0    3     2    0    2    2105.97987   2105.98038   -0.00051
    3    0    3     3    1    2    2104.59845   2104.59849   -0.00004
    3    1    3     3    1    2    2104.74660   2104.74520    0.00139
    3    1    3     2    1    2    2105.92271   2105.92185    0.00085
    3    1    3     2    0    2    2106.12784   2106.12709    0.00074
    3    2    1     2    2    0    2106.00260   2106.00208    0.00051
    3    2    1     3    1    2    2105.76595   2105.76509    0.00085
    3    2    1     3    2    2    2105.09890   2105.09784    0.00105
    3    3    1     3    3    0    2105.08050   2105.07991    0.00058
    3    3    1     2    2    0    2107.34707   2107.34794   -0.00087
    4    0    4     3    1    3    2106.09208   2106.09218   -0.00010
    4    0    4     3    0    3    2106.23729   2106.23796   -0.00067
    4    1    4     3    1    3    2106.18511   2106.18506    0.00004
```

| | | | | | | | | |
|---|---|---|---|---|---|---|---|---|
| 4 | 2 | 2 | 3 | 2 | 1 | 2106.32845 | 2106.32952 | −0.00107 |
| 4 | 2 | 2 | 4 | 2 | 3 | 2105.16416 | 2105.16430 | −0.00014 |
| 4 | 1 | 3 | 3 | 1 | 2 | 2106.41556 | 2106.41529 |  0.00026 |
| 4 | 1 | 3 | 4 | 1 | 4 | 2105.69117 | 2105.68869 |  0.00247 |
| 4 | 2 | 3 | 4 | 2 | 2 | 2104.93971 | 2104.94026 | −0.00055 |
| 4 | 2 | 3 | 3 | 2 | 2 | 2106.26965 | 2106.27036 | −0.00071 |
| 5 | 1 | 4 | 4 | 1 | 3 | 2106.71492 | 2106.71285 |  0.00206 |
| 5 | 2 | 4 | 4 | 2 | 3 | 2106.56175 | 2106.56133 |  0.00041 |
| 5 | 2 | 4 | 5 | 2 | 3 | 2104.81167 | 2104.81105 |  0.00061 |
| 5 | 0 | 5 | 4 | 0 | 4 | 2106.47992 | 2106.48103 | −0.00111 |
| 5 | 0 | 5 | 4 | 1 | 4 | 2106.38836 | 2106.38885 | −0.00049 |
| 5 | 1 | 5 | 4 | 1 | 4 | 2106.44136 | 2106.44243 | −0.00107 |
| 5 | 1 | 5 | 4 | 0 | 4 | 2106.53355 | 2106.53462 | −0.00107 |
| 6 | 0 | 6 | 5 | 0 | 5 | 2106.71940 | 2106.71724 |  0.00215 |
| 6 | 1 | 6 | 5 | 1 | 5 | 2106.69284 | 2106.69322 | −0.00038 |
| 6 | 1 | 5 | 5 | 1 | 4 | 2106.99152 | 2106.99296 | −0.00144 |

Table A2. Observed and calculated transitions in the (1001) <-- (0000) band of He-OCS dimer around 2918 cm$^{-1}$ (units of cm$^{-1}$).

```
**********************************************************************
    J'   Ka'  Kc'    J"   Ka"  Kc"   Observed   Calculated    Obs-Calc
**********************************************************************
    0    0    0     1    1    1    2917.71910  2917.71921   -0.00010
    0    0    0     1    0    1    2917.97583  2917.97537    0.00045
    1    1    0     1    0    1    2918.59374  2918.59340    0.00033
    1    1    0     2    1    1    2917.60706  2917.60711   -0.00004
    1    0    1     1    1    0    2917.96310  2917.96264    0.00045
    1    0    1     2    1    2    2917.47183  2917.47223   -0.00039
    1    0    1     2    0    2    2917.67729  2917.67747   -0.00017
    1    0    1     0    0    0    2918.58467  2918.58448    0.00019
    1    1    1     1    1    0    2918.21812  2918.21746    0.00065
    1    1    1     2    2    0    2916.78766  2916.78737    0.00028
    1    1    1     2    0    2    2917.93286  2917.93228    0.00057
    1    1    1     2    1    2    2917.72709  2917.72704    0.00004
    2    0    2     1    0    1    2918.87574  2918.87610   -0.00036
    2    0    2     2    1    1    2917.88962  2917.88981   -0.00019
    2    0    2     3    0    3    2917.39503  2917.39540   -0.00037
    2    1    2     1    1    1    2918.82403  2918.82422   -0.00018
    2    1    2     3    1    3    2917.45382  2917.45391   -0.00009
    2    1    2     2    1    1    2918.09445  2918.09409    0.00035
    2    1    2     3    0    3    2917.59947  2917.59969   -0.00022
    2    1    2     1    0    1    2919.08050  2919.08039    0.00011
    2    2    0     1    1    1    2919.75876  2919.75870    0.00005
    2    2    0     3    2    1    2917.32048  2917.32075   -0.00027
    2    2    0     2    1    1    2919.02887  2919.02858    0.00029
    2    2    0     2    2    1    2918.27799  2918.27761    0.00037
    2    1    1     2    0    2    2918.65758  2918.65752    0.00005
    2    1    1     2    1    2    2918.45258  2918.45228    0.00030
    2    1    1     3    1    2    2917.27559  2917.27563   -0.00004
    2    1    1     2    2    0    2917.51253  2917.51261   -0.00008
    2    1    1     1    1    0    2918.94218  2918.94270   -0.00051
    2    2    1     2    2    0    2918.25998  2918.25957    0.00040
    2    2    1     3    2    2    2917.35511  2917.35533   -0.00021
    3    1    2     4    1    3    2916.95253  2916.95237    0.00016
    3    1    2     3    2    1    2917.55427  2917.55487   -0.00060
    3    1    2     2    1    1    2919.26270  2919.26269    0.00000
    3    1    2     3    0    3    2918.76811  2918.76829   -0.00017
    3    1    2     3    1    3    2918.62202  2918.62251   -0.00049
    3    2    2     2    2    1    2919.17568  2919.17567    0.00000
    3    2    2     4    2    3    2917.05344  2917.05358   -0.00014
    3    2    2     3    2    1    2918.21812  2918.21881   -0.00069
    3    2    2     2    1    1    2919.92639  2919.92663   -0.00024
    3    3    0     4    3    1    2917.02094  2917.01979    0.00114
    3    3    0     2    2    1    2920.53148  2920.53173   -0.00024
    3    0    3     4    1    4    2917.03959  2917.03940    0.00018
    3    0    3     4    0    4    2917.13138  2917.13158   -0.00019
    3    0    3     3    1    2    2917.76597  2917.76600   -0.00003
    3    0    3     2    0    2    2919.14787  2919.14789   -0.00002
```

| | | | | | | | | |
|---|---|---|---|---|---|---|---|---|
| 3 | 1 | 3 | 3 | 1 | 2 | 2917.91109 | 2917.91124 | -0.00014 |
| 3 | 1 | 3 | 4 | 1 | 4 | 2917.18463 | 2917.18463 | -0.00000 |
| 3 | 1 | 3 | 2 | 1 | 2 | 2919.08776 | 2919.08789 | -0.00012 |
| 3 | 2 | 1 | 3 | 2 | 2 | 2918.30569 | 2918.30554 | 0.00015 |
| 3 | 2 | 1 | 2 | 0 | 2 | 2920.35486 | 2920.35469 | 0.00016 |
| 3 | 2 | 1 | 2 | 1 | 2 | 2920.14927 | 2920.14945 | -0.00017 |
| 3 | 2 | 1 | 3 | 1 | 2 | 2918.97246 | 2918.97280 | -0.00033 |
| 3 | 2 | 1 | 2 | 2 | 0 | 2919.20982 | 2919.20978 | 0.00003 |
| 3 | 2 | 1 | 4 | 2 | 2 | 2916.97513 | 2916.97543 | -0.00030 |
| 3 | 3 | 1 | 2 | 2 | 0 | 2920.52174 | 2920.52184 | -0.00009 |
| 3 | 3 | 1 | 4 | 3 | 2 | 2917.02619 | 2917.02432 | 0.00186 |
| 4 | 0 | 4 | 3 | 1 | 3 | 2919.25196 | 2919.25196 | 0.00000 |
| 4 | 0 | 4 | 3 | 0 | 3 | 2919.39775 | 2919.39773 | 0.00001 |
| 4 | 0 | 4 | 4 | 1 | 3 | 2917.58139 | 2917.58181 | -0.00041 |
| 4 | 1 | 4 | 3 | 1 | 3 | 2919.34350 | 2919.34387 | -0.00036 |
| 4 | 1 | 4 | 3 | 0 | 3 | 2919.48922 | 2919.48964 | -0.00042 |
| 4 | 1 | 4 | 5 | 1 | 5 | 2916.91980 | 2916.91933 | 0.00046 |
| 4 | 2 | 2 | 4 | 2 | 3 | 2918.37430 | 2918.37389 | 0.00041 |
| 4 | 2 | 2 | 3 | 1 | 3 | 2920.60720 | 2920.60675 | 0.00044 |
| 4 | 2 | 2 | 5 | 2 | 3 | 2916.62354 | 2916.62361 | -0.00007 |
| 4 | 2 | 2 | 3 | 2 | 1 | 2919.53903 | 2919.53911 | -0.00007 |
| 4 | 3 | 2 | 3 | 3 | 1 | 2919.47374 | 2919.47497 | -0.00123 |
| 4 | 3 | 2 | 4 | 3 | 1 | 2918.23955 | 2918.23933 | 0.00021 |
| 4 | 3 | 2 | 3 | 2 | 1 | 2920.79294 | 2920.79441 | -0.00146 |
| 4 | 4 | 0 | 4 | 4 | 1 | 2918.23442 | 2918.23447 | -0.00005 |
| 4 | 1 | 3 | 5 | 1 | 4 | 2916.64257 | 2916.64199 | 0.00057 |
| 4 | 1 | 3 | 4 | 0 | 4 | 2918.93435 | 2918.93518 | -0.00082 |
| 4 | 1 | 3 | 3 | 1 | 2 | 2919.56938 | 2919.56959 | -0.00021 |
| 4 | 2 | 3 | 5 | 2 | 4 | 2916.75684 | 2916.75671 | 0.00012 |
| 4 | 2 | 3 | 3 | 2 | 2 | 2919.46236 | 2919.46250 | -0.00013 |
| 4 | 2 | 3 | 3 | 1 | 2 | 2920.12934 | 2920.12976 | -0.00041 |
| 4 | 2 | 3 | 4 | 2 | 2 | 2918.13223 | 2918.13239 | -0.00016 |
| 4 | 3 | 1 | 3 | 3 | 0 | 2919.47886 | 2919.47979 | -0.00092 |
| 4 | 4 | 1 | 4 | 4 | 0 | 2918.23442 | 2918.23437 | 0.00004 |
| 5 | 1 | 4 | 4 | 1 | 3 | 2919.85793 | 2919.85866 | -0.00073 |
| 5 | 1 | 4 | 6 | 1 | 5 | 2916.35100 | 2916.34924 | 0.00176 |
| 5 | 2 | 4 | 4 | 2 | 3 | 2919.73941 | 2919.73956 | -0.00014 |
| 5 | 3 | 2 | 5 | 3 | 3 | 2918.25463 | 2918.25415 | 0.00047 |
| 5 | 3 | 2 | 4 | 3 | 1 | 2919.78420 | 2919.78328 | 0.00092 |
| 5 | 3 | 2 | 4 | 2 | 3 | 2921.17306 | 2921.17313 | -0.00007 |
| 5 | 0 | 5 | 6 | 0 | 6 | 2916.63318 | 2916.63291 | 0.00027 |
| 5 | 0 | 5 | 4 | 0 | 4 | 2919.63102 | 2919.63150 | -0.00048 |
| 5 | 1 | 5 | 4 | 1 | 4 | 2919.59111 | 2919.59195 | -0.00084 |
| 5 | 1 | 5 | 6 | 1 | 6 | 2916.65823 | 2916.65729 | 0.00093 |
| 5 | 2 | 3 | 6 | 2 | 4 | 2916.27750 | 2916.27659 | 0.00091 |
| 5 | 2 | 3 | 4 | 2 | 2 | 2919.86917 | 2919.86908 | 0.00009 |
| 5 | 3 | 3 | 4 | 2 | 2 | 2921.02800 | 2921.03048 | -0.00247 |
| 5 | 3 | 3 | 4 | 3 | 2 | 2919.76732 | 2919.76730 | 0.00002 |
| 6 | 0 | 6 | 7 | 0 | 7 | 2916.38537 | 2916.38382 | 0.00155 |
| 6 | 0 | 6 | 5 | 0 | 5 | 2919.85793 | 2919.85775 | 0.00018 |
| 6 | 1 | 6 | 5 | 1 | 5 | 2919.83159 | 2919.83265 | -0.00106 |
| 6 | 1 | 6 | 7 | 1 | 7 | 2916.39960 | 2916.39727 | 0.00233 |
| 6 | 1 | 6 | 5 | 0 | 5 | 2919.88394 | 2919.88556 | −0.00161 |
| 6 | 2 | 4 | 5 | 2 | 3 | 2920.18867 | 2920.18846 | 0.00020 |
| 6 | 3 | 4 | 5 | 3 | 3 | 2920.05376 | 2920.05302 | 0.00073 |
| 6 | 1 | 5 | 5 | 1 | 4 | 2920.12534 | 2920.12545 | −0.00010 |
| 6 | 2 | 5 | 5 | 2 | 4 | 2920.00599 | 2920.00574 | 0.00025 |
| 6 | 3 | 3 | 5 | 3 | 2 | 2920.09400 | 2920.09212 | 0.00188 |
| 7 | 1 | 6 | 6 | 1 | 5 | 2920.36824 | 2920.36807 | 0.00017 |
| 7 | 2 | 6 | 6 | 2 | 5 | 2920.26151 | 2920.26090 | 0.00060 |
| 7 | 3 | 4 | 6 | 3 | 3 | 2920.40870 | 2920.40792 | 0.00077 |
| 7 | 0 | 7 | 6 | 1 | 6 | 2920.05376 | 2920.05310 | 0.00065 |
| 7 | 1 | 7 | 6 | 1 | 6 | 2920.06592 | 2920.06672 | −0.00080 |
| 7 | 2 | 5 | 6 | 2 | 4 | 2920.48824 | 2920.48998 | −0.00173 |
| 7 | 3 | 5 | 6 | 3 | 4 | 2920.32950 | 2920.33080 | −0.00129 |
| 8 | 0 | 8 | 7 | 0 | 7 | 2920.30258 | 2920.30307 | −0.00048 |
| 8 | 1 | 7 | 7 | 1 | 6 | 2920.59034 | 2920.59011 | 0.00022 |
| 8 | 2 | 7 | 7 | 2 | 6 | 2920.50625 | 2920.50625 | 0.00000 |

Table A3. Observed and calculated transitions in the (0401) <-- (0000) band of He-OCS dimer around 2938 cm$^{-1}$ (units of cm$^{-1}$).

```
************************************************************************
   J'   Ka'  Kc'     J"   Ka"  Kc"   Observed    Calculated    Obs-Calc
************************************************************************
    0    0    0      1    0    1    2937.17651   2937.17754   -0.00102
    1    1    0      2    1    1    2936.81893   2936.81832    0.00061
    1    0    1      0    0    0    2937.78770   2937.78772   -0.00002
    1    0    1      2    0    2    2936.88066   2936.88072   -0.00005
    1    1    1      2    1    2    2936.93922   2936.93859    0.00062
    1    1    1      1    1    0    2937.42894   2937.42901   -0.00006
    2    0    2      1    0    1    2938.08216   2938.08194    0.00022
    2    0    2      3    0    3    2936.60214   2936.60124    0.00089
    2    1    2      1    1    1    2938.03759   2938.03772   -0.00013
    2    1    2      2    1    1    2937.30659   2937.30759   -0.00100
    2    1    2      3    1    3    2936.66705   2936.66741   -0.00035
    2    2    0      2    2    1    2937.46284   2937.46386   -0.00101
    2    1    1      1    1    0    2938.15625   2938.15578    0.00046
    2    2    1      2    2    0    2937.44570   2937.44479    0.00090
    3    1    2      2    1    1    2938.47898   2938.47940   -0.00041
    3    0    3      2    0    2    2938.35840   2938.35818    0.00022
    3    1    3      2    1    2    2938.30508   2938.30434    0.00073
    4    0    4      3    0    3    2938.61401   2938.61420   -0.00019
    4    1    4      3    1    3    2938.56394   2938.56460   -0.00065
    4    2    2      3    2    1    2938.73482   2938.73470    0.00011
    4    1    3      3    1    2    2938.79062   2938.79075   -0.00013
    5    0    5      4    0    4    2938.85581   2938.85592   -0.00010
    5    1    5      4    1    4    2938.81945   2938.81905    0.00040
```